\documentclass[12pt]{iopart}
\usepackage{harvard}
\usepackage{setstack}
\usepackage{iopams}
\usepackage{multirow}
\usepackage{enumitem}
\usepackage{float} 
\usepackage{algorithm}        
\usepackage{algpseudocode}    
\usepackage{amsthm}
\usepackage{array}
\usepackage{geometry}         
\usepackage{soul}             
\usepackage{url}
\usepackage{multirow} 
\usepackage{graphicx}
\usepackage[colorlinks=true, linkcolor=black]{hyperref}
\usepackage[all]{hypcap} 
\usepackage{makecell} 
\usepackage{svg}  
\begin{document}
\newcommand{\defeq}{\overset{\mathrm{def}}{=}}
\renewcommand{\arraystretch}{1.2}
    \title[IPF prediction using deep learning]{Prognostic Model for Idiopathic Pulmonary Fibrosis Using Context-Aware  Sequential-Parallel Hybrid Transformer and Enriched Clinical Information}
    \author{Mahdie Dolatabadi, 
    Shahabedin Nabavi and
    Mohsen Ebrahimi Moghaddam}
    \address{Faculty of Computer Science and Engineering, Shahid Beheshti University, Tehran, Iran}
    \eads{\mailto{m.dolatabadi@alumni.sbu.ac.ir},
          \mailto{\{s\_nabavi,m\_moghadam\}@sbu.ac.ir}}
    \begin{abstract}
        Idiopathic pulmonary fibrosis (IPF) is a progressive disease that irreversibly transforms lung tissue into rigid fibrotic structures, leading to debilitating symptoms such as shortness of breath and chronic fatigue. The heterogeneity and complexity of this disease, particularly regarding its severity and progression rate, have made predicting its future course a complex and challenging task. Besides, traditional diagnostic methods based on clinical evaluations and imaging have limitations in capturing the disease's complexity. Using the Kaggle \emph{Pulmonary Fibrosis Progression} dataset, which includes computed tomography images, and clinical information, the model predicts changes in forced vital capacity (FVC), a key progression indicator. Our method uses a proposed context-aware sequential-parallel hybrid transformer model and clinical information enrichment for its prediction. The proposed method achieved a Laplace Log-Likelihood score of $-6.508$, outperforming prior methods and demonstrating superior predictive capabilities. These results highlight the potential of advanced deep learning techniques to provide more accurate and timely predictions, offering a transformative approach to the diagnosis and management of IPF, with implications for improved patient outcomes and therapeutic advancements. Code Availability: \href{https://github.com/mahdie-dolatabadi/IPF-prediction-using-deep-learning}{GitHub}

    \end{abstract}
    \submitto{Physics in Medicine \& Biology}

    \section{Introduction}
        Multiple organs, such as the liver, heart, kidneys, and lungs, can be affected by fibrosis, which causes a major mortality crisis. Fibrosis arises from the excessive accumulation of collagen and other extracellular matrix (ECM) constituents. Several diseases, including non-alcoholic fatty liver disease, hepatitis viruses, chronic kidney diseases, and pulmonary fibrosis, are closely related to fibrosis~\cite{zhao2022targeting}. According to statistics, approximately 4,968 individuals per 100,000 are affected annually by fibrosis complications~\cite{zhao2022targeting}. Among these patients, those suffering from pulmonary fibrosis comprise a wide range of deaths from pulmonary diseases~\cite{nabavi2021medical}. According to~\cite{barratt2018idiopathic}, the incidence rate of this disease has been reported to range between 2.8 to 9.3 per 100,000 individuals in North America and Europe. This rate escalates to 400 individuals per 100,000 in the population over 65~\cite{martinez2017idiopathic}. The most prevalent yet enigmatic form of pulmonary fibrosis is idiopathic pulmonary fibrosis (IPF)~\cite{noble2012pulmonary}. Based on~\cite{bjoraker1998prognostic}, only 25\% of patients diagnosed with IPF survived five years post-diagnosis, a notably lower survival rate compared to other variants such as idiopathic interstitial and nonspecific interstitial pneumonia. IPF, like all forms of pulmonary fibrosis, exhibits a progressive course with no therapeutic approach currently demonstrating efficacy in halting its advancement. An important aspect to consider alongside the severity of this condition is its escalating prevalence. A UK-based primary care database study reports a 78\% increase in the frequency of this ailment between 2000 and 2012, posing a significant economic burden on global healthcare~\cite{barratt2018idiopathic}. The predominant indicator of this variant is shortness of breath~\cite{nakamura2015idiopathic}. Other symptoms encompass swelling of the fingers (characterized by enlarged and rounded fingertips), chronic dry cough, sudden weight loss, fatigue, and weakness.

        Using artificial intelligence (AI) for tracking the progression of pulmonary fibrosis through demographic data and medical image analysis has grown in recent years. The combination of machine learning models and image analysis techniques can improve timely disease diagnosis~\cite{badrigilan2021deep}. The intricate patterns and features, which are beyond the discernment of the human vision in most cases, can be detected and analyzed by these methodologies. A challenge was raised by the open-source imaging consortium (OSIC)\footnote{https://www.osicild.org/} to predict pulmonary fibrosis progression using AI in 2020. This challenge was organized as a competition hosted by Kaggle~\cite{kaggleOSICPulmonary}, the world's premier platform for scientific communities. The participants were asked to predict lung capacity at certain times using the released dataset in this competition.

        The top performer in the contest attained the highest accuracy through amalgamating EfficientNet B5 architecture with quantile regression dense neural network~\cite{towardsdatascienceAchievedPlace}, leading to subsequent extensive research by numerous scientific groups on this dataset post-competition. In one of these studies~\cite{mandal2020prediction}, a comparative analysis of three regression methods, including ridge, lasso, and elastic-net regression, was conducted and convolutional neural networks (CNNs) were employed as the primary learning framework. In specific cases, researchers have suggested customized models such as depthwise and pointwise convolutions to address task requirements~\cite{wong2021fibrosis}. Considering that pulmonary fibrosis leads to the development of honeycomb patterns in the lungs,~\cite{al2021fibro} attended to the model's need to concentrate on certain regions of the CT images. In addition to focusing on these regions using a self-attention approach, the study recognized the critical need to overcome dataset scarcity via utilizing pre-trained weights within CNNs~\cite{al2021fibro}. Another contentious topic is the type of regression analyzed in~\cite{chutia2021analysis}. Despite all cutting-edge technologies, machine learning algorithms like GBDT and NGBoost have been implemented in~\cite{glotov2021pulmonary} to compare them with state-of-the-art models.

        Previous studies have overlooked key aspects of the challenges involved in this domain. One major limitation is the lack of precision in mask generation, which results in the loss of critical edge information around the lungs. Additionally, existing works predominantly focus on local or global feature extraction without adequately investigating approaches that seamlessly combine these features. Lastly, the significant imbalance between the negligible dimensionality of clinical features and the high-dimensional feature vectors derived from imaging data has been largely neglected.
        
        To address these limitations and optimize the process, the proposed method has the following contributions:

        \begin{itemize}
            \item Proposing a context-aware sequential-parallel hybrid transformer model for image feature extraction that can predict forced vital capacity (FVC) by relying on local and global features of CT images.
            \item Using a clinical information enrichment model to increase the impact of patient medical history in the model's final decision.
            \item Outperforming all previous studies including winners of the OSIC challenge from Kaggle. 
        \end{itemize}

    \section{Materials And Methods}
\begin{figure}[t]
            \centering
            \includegraphics[width=\linewidth]{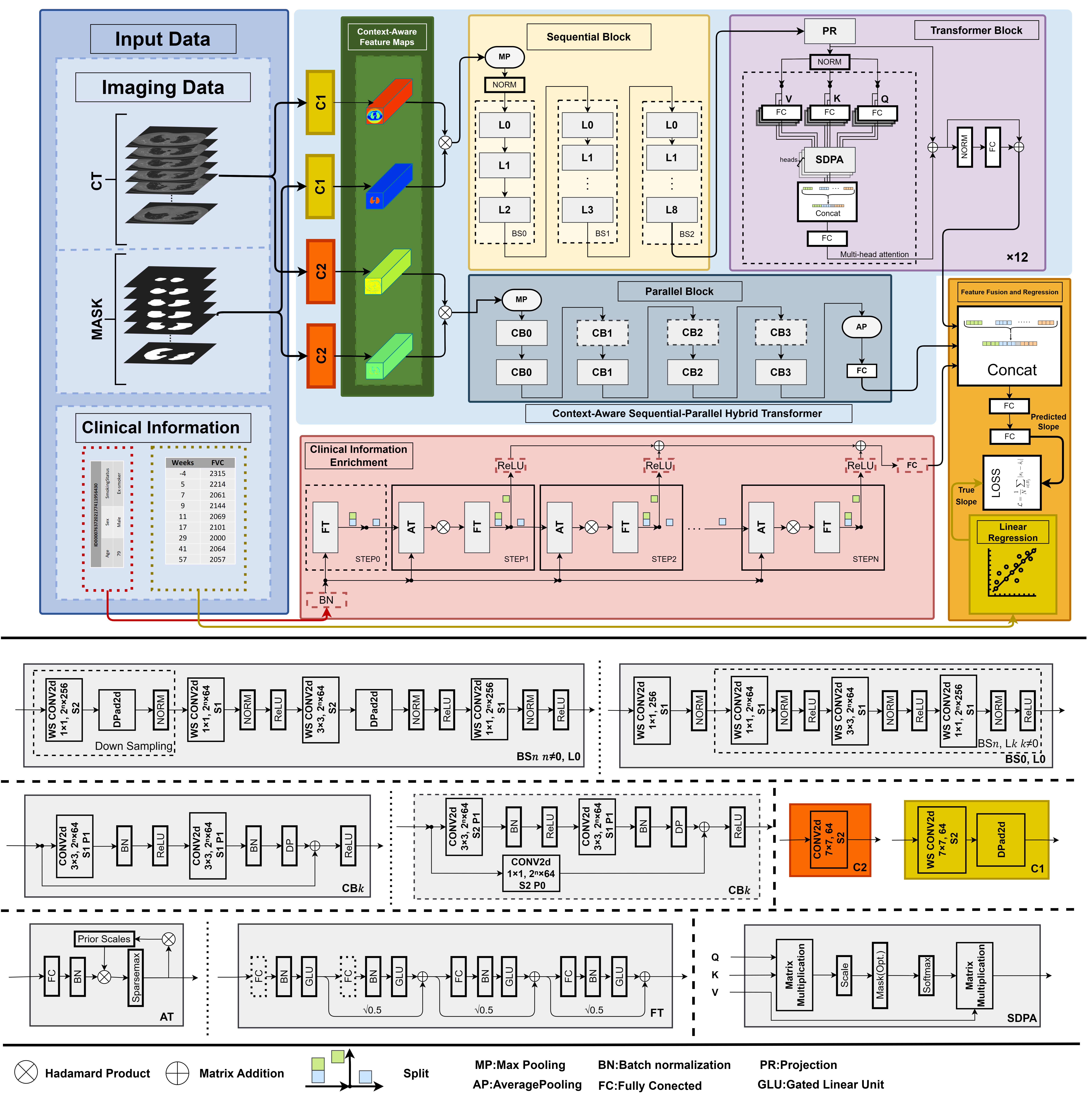}
            \caption{An overview of the proposed method and its main parts.}
            \label{fig:model-arc}
            \end{figure}

        An overview of the proposed method is shown in \autoref{fig:model-arc}. In this proposed method, the data is fed to the proposed deep learning model after data preparation to predict the FVC value for the patient. In this section, the problem is first formulated, and then the different parts of the proposed method are explained.
        
        \subsection{problem formulation}
            Suppose the objective is to address a regression problem focused on predicting pulmonary function over time. In particular, the proposed method is designed to estimate the slope of FVC across multiple time points for individual patients, leveraging imaging data and clinical information. Thus, the input data of the proposed method includes CT images, extracted masks for the lung region, and clinical information. The dataset $\mathcal{D}$ can be expressed as:
            \begin{eqnarray}
                \mathcal{D} = \{(x_i, y_i)\}_{i=1}^{N}
            \end{eqnarray}
            where $x_i$ and $y_i$ indicate the input data and corresponding FVC values at $n$ distinct time points. Thus, $x_i$ and $y_i$ are described based on equations 2 and 3 as follows:
            \begin{eqnarray}
                x_i = \langle I_i, M_i, C_i \rangle
            \end{eqnarray}
            \begin{eqnarray}
                y_i = \{(t_j, m_j) \mid j = 1, 2, \ldots, n\} \quad \nonumber \\
                \hspace{3em} \forall \, j, k \in \{1, 2, \ldots, n\}, \text{ if } t_j \prec t_k, \text{ then } (t_j, m_j) \prec (t_k, m_k)
            \end{eqnarray}
            where $I_i$, $M_i$, and $C_i$ are the input CT stack, the corresponding extracted masks for the lung region, and the clinical information for the $i^{\text{th}}$ patient, respectively. For a set of FVC measurements $y_i$, where $t_j$ is the timestamp and $m_j$ is the FVC measurement value, $(t_j, m_j) \prec (t_k, m_k)$. Here, $\prec$ denotes the precedence relation based on the timestamps. The set $y_i$ is ordered chronologically by their timestamps $t_j$.

        \subsection{The proposed method}
            The proposed method consists of three main parts as follows:
            \begin{enumerate} 
                \item Context-aware sequential-parallel hybrid transformer model for image feature extraction
                \item Clinical information enrichment model for increasing the impact of patient medical history
                \item Feature fusion and regression module
            \end{enumerate}
            The following sections describe each of these parts and explain the proposed method's training and evaluation.

            \subsubsection{Context-aware sequential-parallel hybrid transformer model}
\begin{figure}[htbp]
                \centering
                \includegraphics[width=0.5\textwidth, height=11.5cm]{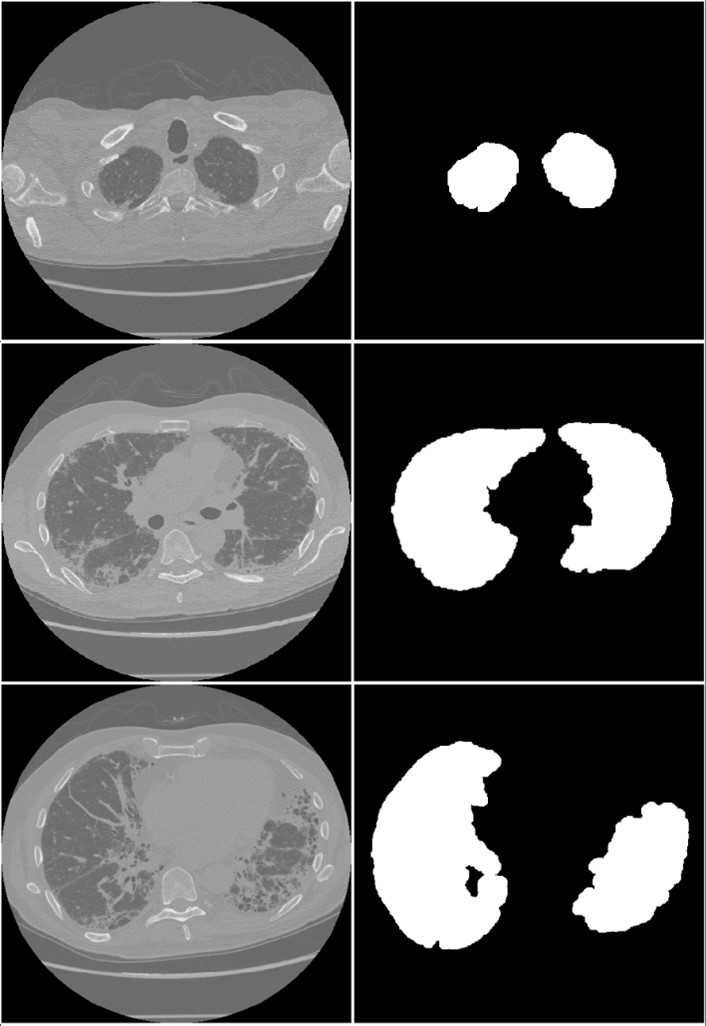}
                \caption{Masks generated by the region growing algorithm for an upper lung slice (first row), middle slice (second row), and lower slice (third row) of the lung region.}
                \label{fig:region-growing}
            \end{figure}

                The mask corresponding to the lung region is first extracted from the input CT images using an automated method to make the context-aware model. We implemented an image processing workflow on the CT imaging dataset to automate the mask generation process and obviate the necessity for radiologist participation. This approach incorporated a region-growing algorithm \cite{gonzalez2009digital}, initiated by defining a seed point within the lung region, followed by morphological operations. Specifically, a dilation operator employing a circular structuring element was utilized after the region-growing segmentation. Examples of masks extracted using this method are shown in \autoref{fig:region-growing}.
                
                The extracted masks should be used to segment CT images to make the model context-aware. The simplest solution to extract ROI in an image is to use the mask product in CT images to obtain the segmented region. These ROIs can then provide the model with only the specific areas to focus on. However, these masks are not always sufficiently accurate, and their direct multiplication with the image may result in the loss of important features. This issue is particularly critical in our work, as most fibrotic regions are located at the edges of the lung lobules in subpleural regions ~\cite{souza2005idiopathic}, and the objective is to predict the severity of pulmonary fibrosis based on these images. Using inaccurate or inappropriate masks can significantly affect the model’s performance. To address this issue, our proposed solution involves multiplying the mask and the image after they pass through a convolutional layer. This approach ensures that if important information is inadvertently removed during direct multiplication, the trainable convolutional layer can process and recover these features, minimizing the risk of losing critical information. \\
                A novel hybrid transformer architecture is leveraged to extract imaging features from the extracted ROIs. In this architecture, a CNN-based model extracts features simultaneously and in parallel with a sequential hybrid model, including a CNN model followed by a vision transformer (ViT). Thus, we called this proposed hybrid model a sequential-parallel hybrid transformer model. CNNs are highly effective at capturing intricate local features but struggle with modeling long-range dependencies and are constrained by their requirement for fixed input sizes. Conversely, ViTs overcome these drawbacks by capturing global contextual relationships and accommodating variable input sizes. By leveraging the complementary strengths, hybrid architectures have emerged as a powerful strategy to address their limitations, providing a balanced and efficient solution for complex image analysis tasks.  \autoref{fig:model-arc} shows the architecture of the context-aware sequential-parallel hybrid transformer model.

            \subsubsection{Clinical information enrichment model}
            Since the importance of some clinical information such as age, gender, and smoking status in the prediction of pulmonary fibrosis has been proven ~\cite{yoon2024smoking,han2008sex}, these data were used as input in the training process of the proposed method. Age in normalized form, gender as values of 0 and 1, and smoking status as a 2-digit binary code are the influential inputs. Thus, the clinical feature vector contains four values. Given that the image feature vector has a much larger dimension than the clinical feature vector, a feature enrichment process was used to enhance and increase the value of clinical features. Many existing methods have not adequately addressed the enhancement and enrichment of tabular demographic and clinical data. However, innovative approaches, such as those discussed in ~\cite{liu2022novel}, have shown promising outcomes. Specifically, as highlighted in ~\cite{arik2021tabnet}, employing the TabNet model for this purpose can be highly effective. Unlike many deep learning models originally developed for image or text data, TabNet is a deep learning model specifically designed for tabular data. TabNet directly processes demographic data, leveraging attention-based transformers to selectively focus on and enhance relevant and valuable features during the learning process.

            \subsubsection{Feature fusion and regression module}
            The feature fusion module integrates a feature vector extracted by the Context-aware sequential-parallel hybrid transformer model and another vector achieved from the clinical information enrichment process. Subsequently, the fused representation is processed through a regression module, which predicts each patient's disease progression rate. These modules estimate the rate of disease progression, based on which FVC is calculated, by concatenating feature vectors and then feeding the concatenated vector to a multilayer perceptron.

            \subsubsection{Model training}
            These three main parts work together as an end-to-end model to effectively process imaging and clinical information for the prediction task. \hyperref[alg:prognostic_model] {Algorithm~\ref{alg:prognostic_model}} outlines the training procedure of the proposed method. 

            \begin{algorithm}[H]
                \caption{Prognostic Model for Idiopathic Pulmonary Fibrosis}
                \label{alg:prognostic_model} 
                \begin{algorithmic}[1]  
                \Require
                    \Statex$\triangleright\ \mathcal{D} = \{(x_i, y_i)\}_{i=1}^{N} = \{ \langle I_i, M_i, C_i \rangle, (t_j, m_j)\}_{i=1, j=1}^{N, N^{'}}$ is an annotated dataset 
                     
                    \Statex$\triangleright\ \theta:$ Initial parameters of the model
                    \Ensure
                     \Statex $\triangleright\ \theta^{'}:$ Final parameters of the model 

                    \For{each fold $k \in \{1, 2, \dots\}$}
                        \For{each epoch $e \in \{1, 2, \dots \}$}
                            \State Select a $batch$ from  $\mathcal{D}^k_{\text{train}}$ 
                                \For{each patient $( \langle I_i, M_i, C_i \rangle, (t_j, m_j)) \in batch $}
                                    \State $\hat{s}_i = f_{\theta} ( I_i, M_i, C_i;  \theta)$
                                    \State $
                                            s = \frac{(n - 1) \sum_{j=1}^{n}{t_jm_j} - \sum_{j=1}^{n}{\sum_{l=1, j\neq l}^{n}{t_lm_j}}}{n\sum_{j=1}^{n}{t_j^{2}} - (\sum_{j=1}^{n}{t_j})^2}
                                            $
                                \EndFor
                                \State $\mathcal{L} = \frac{1}{|batch|} \sum_{i \in batch} | s_i - \hat{s}_i |$
                                \State $\theta^{'} \leftarrow$ AdamW\_Optimizer $(\mathcal{L}, \theta)$
                        \EndFor
                    \EndFor
                    \State \Return $\theta^{'}  $
                \end{algorithmic}
            \end{algorithm}

            \noindent To train $f_{\theta}$, the proposed end-to-end model, it is necessary to provide the data $x_i$ as input to this model:

             \begin{eqnarray}
                f_{\theta}: x_i \rightarrow \hat{s}_i
            \end{eqnarray}
            
            \noindent $\hat{s}_i$ is the prediction of the model based on which the FVC value should be calculated. Thus, $\hat{s}_i$ can be considered the rate of disease progression. 

            \noindent The ground truth value $s_i$ required to train the model is calculated through equation 5, which is the actual value of the disease progression rate.

            \begin{eqnarray}
                m_j = s_i \cdot t_j + m_1, \quad \forall j \in \{1, 2, \ldots, n\}
            \end{eqnarray}

            \noindent To calculate $s_i$, the following relationship, which is in the form of matrix calculations, must be established:

            \begin{eqnarray}
                \mathcal{M}_{i} = \tau_i \beta_i
            \end{eqnarray}

            \noindent where, $\mathcal{M}_{i}$ and $\tau_i$ are defined based on $y_i$, while $\beta_i$ indicates a linear regression parameters. All these entities are matrices, defined as follows:

            \begin{eqnarray}
                \tau_i \triangleq \left[\begin{array}{cc}
                        1 & t_1 \\
                        1 & t_2 \\
                        \vdots & \vdots \\
                        1 & t_n
                        \end{array}\right]&\text{, }\mathcal{M}_{i} \triangleq \left[\begin{array}{c}
                     m_1  \\
                     m_2  \\
                     \vdots  \\
                     m_n  
                \end{array}\right]&\text{, }\beta_i \triangleq  \left[\begin{array}{c}
                     c  \\
                     s 
                \end{array}\right]
            \end{eqnarray}

            \noindent where $s$ and $c$ of $\beta_i$ are the slope and $m_1$, respectively. Thus, according to equations 6 and 7, the value of $s$, which for the $i^{th}$ patient will be the same as the ground truth $s_i$, is unknown in the calculations.

            \noindent Based on ~\cite{math11204311}, the objective of the linear regression method is to minimize the residual sum of squares (RSS). So, we have:

            \begin{eqnarray}
                RSS &=& || \mathcal{M}_{i} - \tau_i \beta_i ||^2 \nonumber \\
                &=& (\mathcal{M}_{i} - \tau_i \beta_i)^T (\mathcal{M}_{i} - \tau_i \beta_i) \nonumber \\
                &=& \mathcal{M}_{i}^{T} \mathcal{M}_{i} - \mathcal{M}_{i}^{T} \tau_i \beta_i + {\beta_i}^T \tau_i^T \tau_i \beta_i
            \end{eqnarray}
            \noindent To minimize RSS  respect to $\beta$, take the gradient with respect to $\beta$ and set it to zero in:

            \begin{eqnarray}
                \frac{\partial RSS}{\partial \beta_i} = -2 \tau_i^T \mathcal{M}_{i} + 2 \tau_i^T \tau_i \beta_i
            \end{eqnarray}
            
            \noindent By setting $\frac{\partial \text{RSS}}{\partial \beta_i} = 0$, we obtain:

            \begin{eqnarray}
                \tau_i^T \mathcal{M}_{i} = \tau_i^T \tau_i \beta_i 
                \Rightarrow 
                \beta_i = (\tau_i^T \tau_i)^{-1} \tau_i^T \mathcal{M}_{i}
            \end{eqnarray}      
        \noindent Here, the gradient of both the linear and quadratic terms is applied. The corresponding formulas are provided in Appendix A. From the calculations in Appendix B, where $\beta_i$ is computed, we can deduce that:

        \begin{eqnarray}
            s = \frac{(n - 1) \sum_{j=1}^{n}{t_jm_j} - \sum_{j=1}^{n}{\sum_{l=1, j\neq l}^{n}{t_lm_j}}}{n\sum_{j=1}^{n}{t_j^{2}} - (\sum_{j=1}^{n}{t_j})^2}
        \end{eqnarray}
        \noindent After calculating the true slope $s_i$ and the predicted slope $\hat{s}_i$, the next step involves quantifying the discrepancy between them to guide the training process. This is achieved by defining the loss function $\mathcal{L}$ as the L1~\cite{hodson2022root} distance between the predicted and true slopes, formulated as:

        \begin{eqnarray}
            \mathcal{L} = \frac{1}{N} \sum_{i \in B_j} | s_i - \hat{s}_i |
        \end{eqnarray}
        where \( N \) represents the size of batch $B_j$, \( \hat{s}_i \) denotes the predicted slope, and \( s_i \) signifies the true slope. The computed loss is then backpropagated through the network, updating the model's parameters to minimize the error in slope prediction.

        The model was trained using an NVIDIA GeForce RTX 3090 graphics card with 24GB of GDDR6X memory and a thermal design power (TDP) of 350W. All runs were performed with a batch size of 8 and using 5-fold cross-validation, such that 80\% of the dataset was used for training and 20\% for testing. To train the proposed method, the AdamW optimizer~\cite{loshchilov2017decoupled} was used with a learning rate of 0.0002.

        \subsubsection{Evaluation Metrics}
\begin{figure}[htbp]
                \centering
                \includegraphics[width=0.9\textwidth, height=11.5cm]{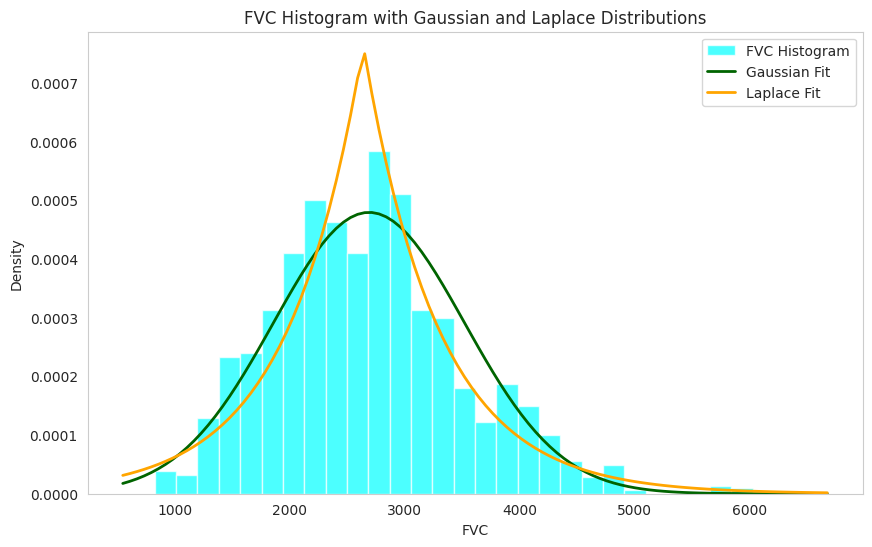}
                \caption{FVC distribution in the dataset and its corresponding Gaussian and Laplacian distributions}
                \label{fig:fvc-distribution}
            \end{figure}

            In medical applications, assessing the reliability and confidence of a model in its predictions is paramount. The quantification of this confidence, often represented as uncertainty, enables clinicians to critically evaluate the model's outputs and make informed decisions regarding their trustworthiness. On the other hand, the choice of evaluation metrics was motivated by the visual resemblance of the FVC dataset distribution to both Gaussian and Laplace distributions as shown in \autoref{fig:fvc-distribution}.
            
            Additionally, it is well-established that Laplace distributions demonstrate greater robustness to noise and outliers compared to Gaussian distributions ~\cite{nair2022maximum}. Consequently, we utilized Laplace log likelihood (LLL) alongside root mean squared error (RMSE)~\cite{hodson2022root} as the evaluation metrics to assess the performance of our regression model. RMSE measures the average magnitude of the prediction errors, defined as the differences between predicted values (\(FVC^{\text{pred}}\)) and actual values (\(FVC^{\text{true}}\)), emphasizing larger errors through the squared term. The formula is as follows:
            \begin{eqnarray}            
            \text{RMSE} = \sqrt{\frac{1}{N} \sum_{i=1}^N (FVC_{i}^{\text{true}} - FVC_{i}^{\text{pred}})^2}
            \end{eqnarray}
            where \(N\) is the total number of samples.
            
            \noindent LLL evaluates the model’s predictions by combining accuracy and uncertainty, assuming a Laplacian distribution for the data. The Laplace distribution is characterized as defined in Equation 14. 
            
            \begin{eqnarray}
            f(x) = \frac{1}{2b} \exp\left(-\frac{|x - \mu|}{b}\right)
            \end{eqnarray}
            where \( \mu \) is the location parameter, and \( b \) is the scale parameter, which defines the shape of the distribution. By substituting $FVC$ into the equation, the Laplace distribution in Equation 14 transforms into:
            \begin{eqnarray}
            f(x) = \frac{1}{2b} \exp\left(-\frac{|FVC^{\text{pred}} - FVC^{\text{true}}|}{b}\right)
            \end{eqnarray}
            Applying the logarithm to the distribution function results in:
            \begin{eqnarray}
            \ln f(x) &= \ln\left(\frac{1}{2b} \exp\left(-\frac{|FVC^{\text{pred}} - FVC^{\text{true}}|}{b}\right)\right) \\
            &= -\ln(2b) - \frac{|FVC^{\text{pred}} - FVC^{\text{true}}|}{b}
            \end{eqnarray}
             The relationship between the scale parameter \( b \) and the variance $\sigma^2$ is established as defined in Equation 18:
            \begin{eqnarray}
            \sigma^2 = 2 b^2
            \end{eqnarray}
            Using Equation 18, the LLL is derived as follows:
            \begin{eqnarray}
            LLL = -\ln(\sqrt{2} \sigma) - \sqrt{2} \frac{|FVC^{\text{pred}} - FVC^{\text{true}}|}{\sigma}
            \end{eqnarray}
            
        \subsection{Data Description}
            The dataset used in this study consists of a limited number of unique identifiers (176 patients), where each patient's data consists of a variable number of CT slices and a tabular dataset containing clinical information and FVC measurements at different timestamps. The \autoref{table:dataset-description} describes the study variables.

        \begin{table}[h!]
            \centering
            \caption{Description of the study variables.}
            \label{table:dataset-description}
            \begin{indented}
            \item[]
            {\scriptsize
            \begin{tabular}{@{}ccccc}
                \br
                \multicolumn{2}{c}{\textbf{Variable}} & \textbf{Frequency(\%)} & \textbf{Range of Values} & \textbf{Mean$\pm$SD} \\ \br
                \multicolumn{5}{|c|}{\textbf{Imaging Data}} \\ \mr
                \multicolumn{2}{c}{\textbf{CT image}}& - & - & - \\ \br
                \multicolumn{5}{|c|}{\textbf{Clinical Information}} \\ \mr
                \multicolumn{2}{c}{\textbf{Age}} & - & 49 - 88 & 67.26$\pm$7.08 \\ \hline
                \multirow{2}{*}{\textbf{Gender}} 
                    & Male & 139 (79\%) & - & -\\ \cline{2-5} 
                    & Female & 37 (21\%) & - & - \\ \mr
                \multirow{3}{*}{\textbf{Smoking Status}} 
                    & Currently smokes & 9 (5\%) & - & -\\ 
                    \cline{2-5}
                    & Ex-smoker & 118 (67\%) & - & -\\ \cline{2-5}
                    & Never smoked & 49 (28\%) & - & -\\ \mr
                \multicolumn{2}{c}{\textbf{FVC}} & 6 - 10 & 827 - 6399 & 2690.47$\pm$832.77 \\ \br

            \end{tabular}
            }
        \end{indented}
        \end{table}

    \section{Experimental Results and Discussion}
         The study results include the proposed method's results, ablation studies, and comparisons with related works. The ablation studies examine the impact of the proposed method's main parts on the final performance. The proposed model's results are also compared with previous studies to show the superiority of the proposed method's performance.

        \subsection{Results of the proposed method and ablation studies}

\begin{figure}
                \centering
                \includegraphics[width=0.7\linewidth]{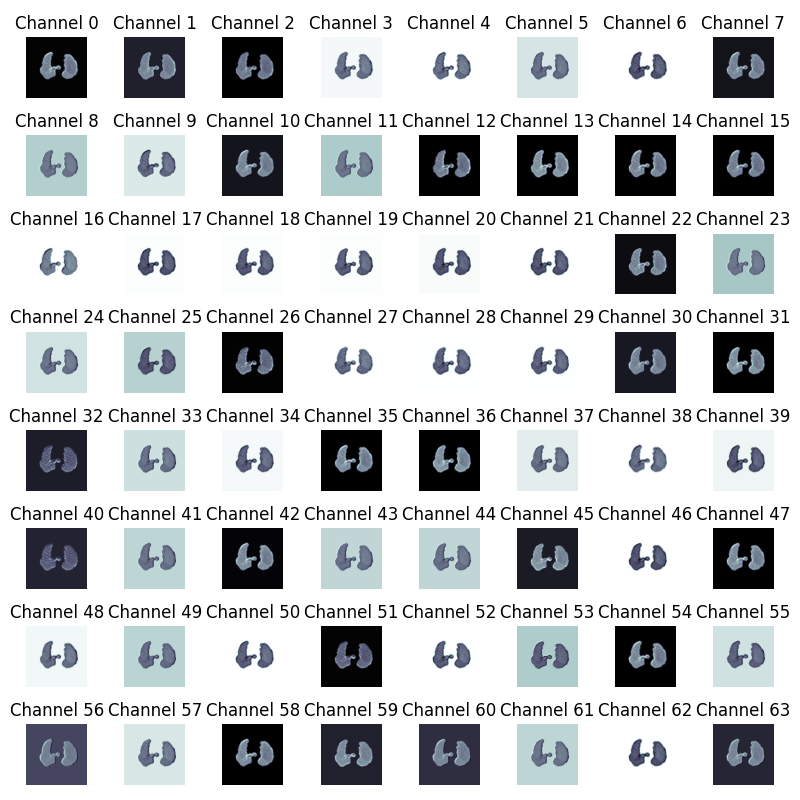}
                \caption{An example of the context-aware module's output}
                \label{fig:context-aware}
            \end{figure}

            The proposed method is made up of various elements and learning strategies, and it seems essential to check the impact of every single element on the final model's performance. The overall method's performance is \textbf{-6.508} for the LLL metric and root mean square error (RMSE) of \textbf{168.26}. However, removing any of the model's elements, including the context-aware module, the clinical information enrichment module, the parallel CNN-based model of the sequential-parallel hybrid transformer model, the sequential part of the sequential-parallel hybrid transformer model, and the transfer learning strategy, results reduction in the value of metrics. This decrease highlights the importance and impact of each element utilized in this study. Results of these ablation studies and the final method performance are presented in \autoref{table:ablation_study}. In addition, the visual representations generated by the context-aware module are also shown in \autoref{fig:context-aware}.

        \begin{table}[h!]
            \centering
            \caption{Results of the proposed method and ablation studies}
            \label{table:ablation_study} 
            \begin{indented}
            \item[]
            {\scriptsize
                \begin{tabular}{@{}ccc}
                    \br
                    \textbf{Experiment} & \textbf{LLL} & \textbf{RMSE} \\ 
                    \mr
                    The proposed method excluding the context-aware module & $-6.55 \pm 0.48$ & $175.10 \pm 44.33$ \\ 
                    The proposed method excluding clinical information enrichment & $-6.54 \pm 0.25$ & $163.37 \pm 38.17$ \\
                    \makecell{The proposed method excluding the sequential part of \\ the sequential-parallel hybrid transformer model} & $-6.54 \pm 0.29$ & $170.85 \pm 41.63$ \\ 
                    The proposed model without transfer learning & $-6.53 \pm 0.25$ & $174.64 \pm 46.15$ \\ 
                    \makecell{The proposed method excluding the parallel CNN-based model of \\ the sequential-parallel hybrid transformer model} & $-6.51 \pm 0.27$ & $169.49 \pm 38.18$ \\ 
                    \br
                    \textbf{The Proposed Method} & {\boldmath $-6.50 \pm 0.21$} & {\boldmath $168.26 \pm 37.74$} \\ 
                    \br
                \end{tabular}
            }
        \end{indented}
        \end{table}
        
        \subsection{Comparison with other related studies}
        \autoref{table:comparison} shows the results of comparing the proposed method with related studies. The proposed method outperforms all previously proposed approaches, including the winner of the Kaggle competition \cite{towardsdatascienceAchievedPlace} and the state-of-the-art methods including ~\cite{al2021fibro} and ~\cite{shehab2023accurate}.

            \begin{table}[h!]
            \centering
            \label{table:comparison}
            \caption{Comparison of the proposed method and previous approaches}
            \begin{indented}
            \item[]
            {\footnotesize 
            \begin{tabular}{@{}cc}
            \br
            \textbf{Approach} & \textbf{LLL} \\ \mr
            ElasticNet ~\cite{glotov2021pulmonary} & $-7.9096$ \\ 
            NGBoost ~\cite{glotov2021pulmonary} & $-6.9362$ \\ 
            Multiple Quantile Regression ~\cite{mandal2020prediction} & $-6.92$ \\ 
            DNN ~\cite{glotov2021pulmonary} & $-6.8842$ \\ 
            GBDT ~\cite{glotov2021pulmonary} & $-6.8826$ \\ 
            Quantile Regression + EfficientNet b5 ~\cite{towardsdatascienceAchievedPlace} & $-6.8305$ \\ 
            FibrosisNet ~\cite{wong2021fibrosis} & $-6.8188$ \\ 
            Ridge Regression \cite{mandal2020prediction} & $-6.81$ \\ 
            ElasticNet Regression ~\cite{mandal2020prediction} & $-6.737$ \\ 
            FibroCosaNet ~\cite{al2021fibro} & $-6.68$ \\ 
            EfficientNet b5 + Noise Addition ~\cite{shehab2023accurate} & $-6.64$ \\ \br
            \textbf{The Proposed Method} & \boldmath $-6.508$ \\ 
            \br
            \end{tabular}
            }
            \end{indented}
            \end{table}

        \subsection{Discussion}
            In this study, an important challenge in medicine has been addressed. The focus is on IPF, an incurable and progressive disease where accurately predicting the rate of its progression is a critical issue. Traditional methods have proved to be inadequate for the purpose. Therefore, there was a need to propose an innovative approach to tackle this challenge. In response, the OSIC organization released a dataset containing imaging and clinical information. Based on this dataset, we tried to propose a new method to make more accurate predictions of the course of this disease. Our proposed method includes providing a context-aware sequential-parallel hybrid transformer model for image feature extraction. This model can simultaneously extract local and global features from CT images by focusing on the lung region. Given that the features extracted from medical images vastly outnumber those of clinical information, the imbalance poses the risk of undermining the respective information's clinical significance within the model. To counter such imbalance, a custom model was employed to better enrich clinical information. The results of this study demonstrate a dramatic improvement in predicting the progression of IPF. Using imaging data and clinical information within a deep learning model enhanced the model's ability to understand the complex patterns of this disease.

            Based on ablation studies, the effectiveness of the various modules of the proposed method was proven, which shows the efficiency of each part of the model in its final performance. Simultaneous extraction of local and global features from CT images by a proposed sequential-parallel hybrid transformer model relying on a context-aware approach and clinical information enrichment has been able to improve the efficiency of the model. Besides, the results of the proposed method clearly show its superior performance compared to other methods, even to the first place in the Kaggle competition. Considering the statistics presented regarding IPF and the necessity of its prediction, it can be said that the proposed method can lead to an improved diagnostic and prognostic process with appropriate accuracy.

            A significant challenge, alongside issues related to optimal model training, is the lack of data for model training and the presence of some slices in CT imaging volumes with irrelevant information. The proposed method is designed to process a single slice as input at each step. This approach enables the utilization of a selected subset of slices containing relevant and meaningful information, alongside the corresponding clinical information associated with each patient. Previous methods often excluded fixed percentages of upper and lower slices in CT scans~\cite{al2021fibro}; however, such approaches carried the risk of discarding valuable information
            due to variations in lung positioning across patients. To mitigate this issue, each patient's CT scan was manually reviewed by a radiologist to identify and record the uppermost and lowermost slices containing relevant information. Ultimately, by presenting a comprehensive and accurate framework, the proposed method not only demonstrated outstanding performance in predicting the progression of IPF but also introduced a novel pathway for leveraging deep learning methods in medical image analysis.

    \section{Conclusion}
        In this study, an end-to-end multimodal model, tailored for a regression task, was developed. To ensure a context-aware learning process, we employed a region-growing algorithm with pre-defined seed points within the lung area to extract lung masks from CT images. In addition, by merging the ViTs and CNNs in a hybrid arrangement, we effectively addressed the limitations of each approach. A clinical feature enrichment mechanism was presented to enhance the representation of clinical information. The quantitative evaluation of this model, including ablation studies and comparisons with prior studies, highlighted its effectiveness and contribution. We will focus on employing three-dimensional models capable of capturing volumetric changes to enhance accuracy by identifying fibrotic regions and extracting complex features with higher precision, as well as incorporating the impact of IPF on lung mechanics through the development of physics-informed neural networks to improve model performance, for our future work.
        
    \appendix
    \section{Gradients of Linear and Quadratic Terms}

            \subsection{Gradient of a Linear Term}
            Consider a function $f(w)$ of the form:
            \begin{eqnarray}
                f(w) = b^\top w,
            \end{eqnarray}
            where $b \in \mathbb{R}^n$ is a vector and $w \in \mathbb{R}^n$ is the variable. The gradient of $f(w)$ with respect to $w$ is:
            \begin{eqnarray}
                \nabla_w f = b.
            \end{eqnarray}

            \noindent This result follows directly from the linearity of $f(w)$ and the fact that the derivative of a linear term with respect to its variable yields the coefficient vector.
            
            \subsection{Gradient of a Quadratic Term}
            Now consider a function $f(w)$ of the form:
            \begin{eqnarray}
            f(w) = w^\top A w,
            \end{eqnarray}
            where $A \in \mathbb{R}^{n \times n}$ is a symmetric matrix ($A = A^\top$), and $w \in \mathbb{R}^n$ is the variable. The gradient of $f(w)$ with respect to $w$ is given by:
            \begin{eqnarray}
            \nabla_w f = 2A w.
            \end{eqnarray}
            \noindent This result can be derived by observing that the quadratic term $w^\top A w$ expands as a scalar product, and due to the symmetry of $A$, the factor of $2$ arises when differentiating with respect to $w$.
                       
    \section{Calculating the slope}
        We start with the following matrices:
        \begin{eqnarray}
            X = \left[\begin{array}{cc}
                    1 & x_1 \\
                    1 & x_2 \\
                    \vdots & \vdots \\
                    1 & x_n
                    \end{array}\right]&\text{, }Y = \left[\begin{array}{cc}
                 y_1  \\
                 y_2  \\
                 \vdots  \\
                 y_n  
            \end{array}\right]&\text{, }\beta = \left[\begin{array}{c} 
                 c  \\
                 m 
            \end{array}\right]
        \end{eqnarray}
        The transpose and multiplication of the matrix $X$ yields:
        \begin{eqnarray}
                X^T X & = & \left[\begin{array}{cccc}
                                1 & 1 & \cdots & 1 \\
                                x_1 & x_2 & \cdots & x_n 
                            \end{array}\right] 
                              .   
                            \left[\begin{array}{cc}
                            1 & x_1 \\
                            1 & x_2 \\
                            \vdots & \vdots \\
                            1 & x_n
                            \end{array}\right] \nonumber\\    
                        & = & \left[\begin{array}{cc}
                                n & x_1 + x_2 + \cdots + x_n \\
                                x_1 + x_2 + \cdots + x_n & x_1^2 + x_2^2 + \cdots + x_n^2                      
                              \end{array}\right]
        \end{eqnarray}
        The inverse of $X^\top X$ is computed as:
        \begin{eqnarray} 
                (X^TX)^{-1} & = & \frac{1}{\underbrace{n(x_1^2 + x_2^2 + \cdots + x_n^2) - (x_1 + x_2 + \cdots + x_n)^2}_{\gamma}} \times \nonumber\\ 
                &&\left[\begin{array}{cc}
                            x_1^2 + x_2^2 + \cdots + x_n^2 & -x_1 - x_2 + \cdots - x_n \\
                            -x_1 - x_2 + \cdots - x_n & n                      \end{array}\right] \triangleq \alpha
        \end{eqnarray}
        Now, consider:
        \begin{eqnarray}
            (X^TX)^{-1}X^T & = & \nonumber\\
            \qquad \alpha X^T & = & \frac{1}{\gamma} \times 
                        \left[
                            \begin{array}{cccc}
                             z_{11} & z_{12} & \cdots & z_{1n} \\
                             z_{21} & z_{22} & \cdots &  z_{2n}
                            \end{array}
                        \right]
        \end{eqnarray}
        \[ \text{where:}
        \left\{ \begin{array}{l} z_{1i} = x_1^2 + x_2^2 + \cdots + x_n^2 - x_i (x_1 + x_2 + \cdots + x_n) \\ 
                                 z_{2i} = - (x_1 + x_2 + \cdots + x_n) + nx_i\\ \end{array} \right. \forall i \in \{1, 2, \ldots, n\}
        \]
        The regression coefficients $\beta$ are computed as:
        \begin{eqnarray}
            \quad \beta & = & \alpha X^T Y \nonumber \\
             & = &  \frac{1}{\gamma} \times \left[\begin{array}{c}
                                     z_{11}y_1 + z_{12}y_2 + \cdots + z_{1n}y_n  \\
                                     z_{21}y_1 + z_{22}y_2 + \cdots + z_{2n}y_n
                                \end{array}\right] \nonumber \\
                           & = & \frac{1}{\gamma} \times \left[\begin{array}{c}
                                 \sum_{i=1}^{n} z_{1i}y_i \\
                                 \sum_{i=1}^{n} z_{2i}y_i 
                            \end{array}\right] =\left[\begin{array}{c}
                                     c  \\
                                     m 
                                \end{array}\right]
        \end{eqnarray}
        The slope $m$ is given by:

        \begin{eqnarray}
            s = \frac{n \sum_{j=1}^{n}{x_jy_j} - \sum_{j=1}^{n}{x_jy_j} - \sum_{j=1}^{n}{\sum_{l=1, j\neq l}^{n}{x_ly_j}}}{n\sum_{j=1}^{n}{x_j^{2}} - (\sum_{j=1}^{n}{x_j})^2}
        \end{eqnarray}
        By factoring out $\sum_{j=1}^{n}{x_jy_j}$, we obtain:
        \begin{eqnarray}
            s = \frac{(n - 1) \sum_{j=1}^{n}{x_jy_j} - \sum_{j=1}^{n}{\sum_{l=1, j\neq l}^{n}{x_ly_j}}}{n\sum_{j=1}^{n}{x_j^{2}} - (\sum_{j=1}^{n}{x_j})^2}
        \end{eqnarray}
    \section*{References}
    \bibliographystyle{jphysicsB}

\end{document}